\documentclass{elsart}

\usepackage{amssymb}
\usepackage[reqno]{amsmath}

\begin{document}

\begin{frontmatter}




\title{Casimir Energy for a Dielectric Cylinder}


\author{In\'es Cavero-Pel\'aez} and
\ead{cavero@nhn.ou.edu}
\author{Kimball A. Milton}
\ead{milton@nhn.ou.edu}
\address{Oklahoma Center for High Energy Physics and Department of
Physics and Astronomy, University of Oklahoma, Norman 73019 USA}

\begin{abstract}
 In this paper we calculate the Casimir energy for a dielectric-diamagnetic
 cylinder with the speed of light differing on the inside and outside.
 Although the result is in general divergent, special cases are meaningful.
 The well-known results for a uniform speed of light are reproduced.  
The self-stress on a purely dielectric cylinder is shown to
 vanish through second order in the deviation of the permittivity from
 its vacuum value, in agreement with the result calculated from the sum
 of van der Waals forces.  These results are unambiguously separated from
 divergent terms.
\end{abstract}

\end{frontmatter}

\date{\today}

 \section{Introduction}
Interest in quantum vacuum phenomena, subsumed under the general
rubric of the Casimir effect, is increasing at a rapid pace.  Status
of work in the field is summarized in recent review articles and monographs
\cite{casrev,casmono,bordagrev}.  
The theoretical developments have been largely
driven by experimental and technological developments, where it is
becoming evident that Casimir forces may present fundamental limits
and opportunities in nanomechanical devices \cite{belllabs} and
nanoelectronics \cite{slinkman}.  Thus it is imperative to understand
fundamental aspects of the theory, such as the sign of the effect,
which, at present,
 cannot be predicted without a detailed calculation.  This paper
represents an incremental increase in our list of solved examples of
Casimir energies with nontrivial boundaries.

The Casimir energy for an uniform dielectric sphere was first calculated in 
1979 by Milton \cite{milton1980}  and later generalized to the case when both 
the electric permittivity and the magnetic permeability are present 
\cite{miltonng97}.  It was 
later observed \cite{ddsph} that, in the special dilute
dielectric case where $\mu=1$ and $|\varepsilon-1|\ll1$, 
the series expansion in $\varepsilon-1$ has a leading term that perfectly 
matches the ``renormalized'' energy obtained by summing the van der Waals 
interactions \cite{milton1998}.  That result
\begin{equation}
E_{\rm vdW}=\frac{23}{1536\pi a}(\varepsilon-1)^2,
\label{vdw}
\end{equation}
is obtained either by isolating and extracting surface and volume
divergences, or directly by analytically continuing in the number of
space dimensions.

The Casimir analysis for the case of the circular cylinder has been attempted on 
several occasions; however, the difficulty of the geometry and the fact that 
the TE and TM modes do not decouple makes the problem considerably more 
complex.  Only in the case when the speeds of light inside and outside
the cylinder coincide is the result completely unambiguous 
\cite{milt-nest-nest,brevik-nyland,Gosdz-romeo,klich-romeo}.  
This includes the classic case
of a perfectly conducting cylindrical shell \cite{deraad&milton1981}
where the energy per unit length is found to be
\begin{equation}
\mathcal{E}=-\frac{0.01356}{a^2},\label{1.1}
\end{equation}
where $a$ is the radius of the cylinder. 
The minus sign indicates that
the Casimir self-stress is attractive, unlike the Boyer repulsion for a sphere
\cite{boyer}.

When the speed of light is different inside and outside of the body, the
Casimir energy will be divergent \cite{milton1980},
which goes beyond those divergences associated with
curvature \cite{Candelas,symanzik,bkv,jaffe,fulling}.  
Thus it seems impossible to ascribe
any significance to results of such calculations.  Any success in extracting
a meaningful result in such cases, as in the dilute dielectric sphere
example mentioned above, seems noteworthy.

We present here the calculation of the Casimir pressure on the walls of
an infinite circular dielectric-diamagnetic cylinder
with electric permittivity $\epsilon$ and magnetic permeability  $\mu$ 
inside the cylinder which is surrounded by vacuum with permittivity
$1$ and permeability $1$ so \mbox{$\varepsilon
\mu \neq1$}. It is shown that the  corresponding Casimir energy per unit length
is divergent, as expected, but, for $\mu=1$, the finite coefficient of
$(\varepsilon-1)^2/a^2$ in the expansion for the dilute approximation yields
the surprising zero result found by summing
the van der Waals energies between the molecules 
that make up the material, in a manner similar to
that which resulted in (\ref{vdw}) \cite{milt-nest-nest,romeopc}.
The latter calculation was independently carried out by Milonni \cite{milonni},
and verified by a perturbative calculation by Barton \cite{barton}.
Although there should be divergences in the energy proportional to
$(\varepsilon-1)^2a$ and $(\varepsilon-1)^2/a$, the coefficient of
$(\varepsilon-1)^2/a^2$ is unique and finite \cite{bordag01}.

The paper is laid out as follows. In Sec.~\ref{sec2} 
we calculate the dyadic Green's 
functions that will allow us to compute the one-loop vacuum expectation values 
of the quadratic field products.  This enables us to calculate the
vacuum expectation value of the stress tensor, the discontinuity of which
across the surface gives the stress on the cylinder, computed in 
Sec.~\ref{sec3}.  The bulk Casimir stress, which would be present if
either medium filled all space, is computed in Sec.~\ref{sec4} and must
be subtracted from the stress found in Sec.~\ref{sec3}.  Finally, the
case of a dilute dielectric cylinder is considered in Sec.~\ref{sec5},
and by detailed analytic and numerical calculations in Sec.~\ref{sec6}
 it is shown that the Casimir stress vanishes both in order
$\varepsilon-1$ and $(\varepsilon-1)^2$.  The significance
of divergences encountered in the calculation is discussed.
The implications of these results
are briefly considered in the Conclusions.

\section{Green's Function Derivation of the Casimir Energy}
\label{sec2}
In order to write down the Green's dyadic equations, we introduce a 
polarization source $\mathbf{P}$ whose linear relation with the electric field 
defines the Green's dyadic as
\begin{equation}
\mathbf{E}(x)=\int (\d x')
\mathbf{\Gamma}(x,x')\cdot \mathbf{P}(x').
\end{equation}
Since the response is translationally invariant in time, we introduce 
the Fourier transform at a given frequency $\omega$,
\begin{equation}
\mathbf{\Gamma}(x,x')=\int_{-\infty}^{\infty}\frac{\d\omega}{2\pi}
\exp{[-\mathrm{i}\omega(t-t')]}\mathbf{\Gamma(r,r'},\omega).
\end{equation}
We can then write the  dyadic Maxwell's equations in a medium characterized by 
a dielectric constant $\varepsilon$ and a permeability $\mu$,
both of which may be functions of frequency  (see 
Ref.~\cite{milton1980,mildersch,schdermil}):
\begin{subequations}
\begin{eqnarray}
\boldsymbol{\nabla}\times\mathbf{\Gamma'}-\mathrm{i}\omega\mu\mathbf{\Phi}
&=&\frac{1}{\varepsilon}\boldsymbol{\nabla}\times\mathbf{1},\label{maxeq1}\\
-\boldsymbol{\nabla}\times\mathbf{\Phi}-\mathrm{i}
\omega\varepsilon\mathbf{\Gamma'}&=&\mathbf{0},
\label{maxeq2}
\end{eqnarray}
\end{subequations}
where
\begin{equation}
\mathbf{\Gamma'(r,r'},\omega)=\mathbf{\Gamma}(\mathbf{r,r'},\omega)
+\frac{\mathbf{1}}{\varepsilon(\omega)},
\end{equation}
and where the unit dyadic $\mathbf{1}$ includes a three-dimensional $\delta$ 
function,
\begin{equation}
\mathbf{1}=\mathbf{1}\delta(\mathbf{r-r'}).
\end{equation}
The two dyadics are solenoidal,
\begin{subequations}
\begin{eqnarray}
\boldsymbol{\nabla\cdot\Phi}&=&\mathbf{0},\\
\boldsymbol{\nabla\cdot\Gamma'}&=&\mathbf{0}.
\end{eqnarray}
\end{subequations}
The corresponding second order equations are
\begin{subequations}
\begin{eqnarray}
(\nabla^2+\omega^2\varepsilon\mu)\mathbf{\Gamma'}&=&
-\frac{\mathbf{1}}{\varepsilon}\boldsymbol{\nabla}\times(
\boldsymbol{\nabla}\times\mathbf{1}),\label{helmholtz1}\\
 (\nabla^2+\omega^2\varepsilon\mu)\mathbf{\Phi}&=&\mathrm{i}
 \omega\boldsymbol{\nabla}
\times\mathbf{1}
\label{helmholtz2}.
\end{eqnarray}
\end{subequations}
Quantum mechanically, these Green's dyadics give the one-loop vacuum 
expectation values of the product of fields at a given frequency $\omega$,
\begin{subequations}
\begin{eqnarray}
\langle\mathbf{E}\mathbf{(r)}\mathbf{E}\mathbf{(r')}\rangle
&=&\frac{\hbar}{\mathrm{i}}\mathbf{\Gamma(r,r')},\label{vevE}\\
\langle\mathbf{H}\mathbf{(r)}\mathbf{H}\mathbf{(r')}\rangle
&=&-\frac{\hbar}{\mathrm{i}}\frac{1}{\omega^2\mu^2}
\overrightarrow{\boldsymbol{\nabla}}
\times\boldsymbol{\Gamma}(\mathbf{r,r'})\times
\overleftarrow{\boldsymbol{\nabla'}}.\label{vevH}
\end{eqnarray}
\end{subequations}
Thus, from the knowledge of the classical Green's dyadics, we can calculate 
the vacuum energy or stress.

We now introduce the appropriate partial wave decomposition for a cylinder,
in terms of cylindrical coordinates $(r,\theta,z)$,  a 
slight modification of that given for a conducting cylindrical shell 
\cite{deraad&milton1981}\footnote{It might be thought that we could immediately
use the general waveguide decomposition of modes into those of TE and
TM type, for example as given in Ref.~\cite{schw&milton2004}.  However, this is
here impossible because the TE and TM modes do not separate.  See 
Ref.~\cite{stratton}.}:
\begin{subequations}
\begin{eqnarray}
\mathbf{\Gamma'}(\mathbf{r,r'};\omega)&=&\sum_{m=-\infty}^{\infty}
\int_{-\infty}^{\infty}\frac{\d k}{2\pi}\bigg{\{}(\boldsymbol{\nabla}\times
\mathbf{\hat z})f_{m}(r;k,\omega)\chi_{mk}(\theta,z)
\nonumber\\
&&\qquad\qquad\mbox{}+\frac{\mathrm{i}}{\omega\varepsilon}\boldsymbol{\nabla}
\times(\boldsymbol{\nabla}\times\mathbf{\hat 
z)}g_m(r;k,\omega)\chi_{mk}(\theta,z)\bigg{\}},\\
\mathbf{\Phi}(\mathbf{r,r'};\omega)&=&\sum_{m=-\infty}^{\infty}
\int_{-\infty}^{\infty}\frac{\d k}{2\pi}\bigg{\{}
(\boldsymbol{\nabla}\times\mathbf{\hat z})\tilde g_{m}(r;k,\omega)
\chi_{mk}(\theta,z)
\nonumber\\
&&\qquad\qquad\mbox{}-\frac{\mathrm{i}\varepsilon}{\omega\mu}\boldsymbol{\nabla}
\times(\boldsymbol{\nabla}\times\mathbf{\hat 
z)}\tilde f_m(r;k,\omega)\chi_{mk}(\theta,z)\bigg{\}},
\end{eqnarray}
\end{subequations}
where the cylindrical harmonics are
\begin{equation}
\chi(\theta,z)=\frac{1}{\sqrt{2\pi}}\e^{\mathrm{i}m\theta}\e^{\mathrm{i}kz},
\end{equation}
and the dependence of $f_m$ etc.\ on $\mathbf{r'}$ is implicit. Notice that 
these are vectors in the second tensor index. Because of the 
presence of these harmonics we have
\begin{subequations}
\begin{eqnarray}
\boldsymbol{\nabla}\times\mathbf{\hat z}&\rightarrow&\mathbf{\hat r}
\frac{\mathrm{i}m}{r}-\boldsymbol{\hat 
\theta}\frac{\partial}{\partial r}\equiv\boldsymbol{\mathcal{M}},\\
\boldsymbol{\nabla}\times(\boldsymbol{\nabla}\times\mathbf{\hat z})
&\rightarrow&\mathbf{\hat 
r}\mathrm{i}k\frac{\partial}{\partial r}-\boldsymbol{\hat \theta}\frac{mk}{r}
-\mathbf{\hat 
z}d_m\equiv\boldsymbol{\mathcal{N}},
\end{eqnarray}
\end{subequations}
in terms of the cylinder operator
\begin{equation}
d_m=\frac{1}{r}\frac{\partial}{\partial r}r\frac{\partial}{\partial 
r}-\frac{m^2}{r^2}.
\end{equation}
Now if we use the Maxwell equation (\ref{maxeq2}) we conclude\footnote{The 
ambiguity in solving for these equations is absorbed in the definition of 
subsequent constants of integration.}
\begin{subequations}
\begin{eqnarray}
\tilde g_m&=&g_m,\\
(d_m-k^2)\tilde f_m&=&-\omega^2\mu f_m.
\end{eqnarray}
\end{subequations}

From the other Maxwell equation (\ref{maxeq1}), we deduce (we now make the 
second, previously suppressed, position arguments explicit; the prime on the 
differential operator signifies action on the second primed argument)
\begin{subequations}
\begin{eqnarray}
d_m\mathcal{D}_m\mathbf{\tilde 
f}_m(r;r',\theta',z')&=&\frac{\omega^2\mu}{\varepsilon}
\boldsymbol{\mathcal{M}'^*}\frac{1}{r}\delta(r-r')\chi_{mk}^*(\theta',z'),\\
d_m\mathcal{D}_m\mathbf{g}_m(r;r',\theta',z')&=&
-\mathrm{i}\omega\boldsymbol{\mathcal{N}'^*}\frac{1}{r}\delta(r-r')
\chi_{mk}^*(\theta',z'),
\end{eqnarray}
\end{subequations}
where the Bessel operator appears,
\begin{equation}
\mathcal{D}_m=d_m+\lambda^2,\qquad\lambda^2=\omega^2\varepsilon\mu-k^2.
\label{besselop}
\end{equation}
Now we separate variables in the second argument,\footnote{Note that here and
in the following there are slight changes in notation, and numerous corrections,
to the treatment sketched in Ref.~\cite{casrev}.}
\begin{subequations}
\begin{eqnarray}
\mathbf{\tilde 
f}_m(r,\mathbf{r'})&=&\left[\boldsymbol{\mathcal{M}'^*}F_m(r,r';k,\omega)
+\frac{1}{\omega}\boldsymbol{\mathcal{N}'^*}\tilde 
F_m(r,r';k,\omega)\right]\chi_{mk}^*(\theta',z'),
\label{ftilde}\\
\mathbf{g}_m(r,\mathbf{r'})&=&\left[-\frac{\mathrm{i}}{\omega}
\boldsymbol{\mathcal{N}'^*}G_m(r,r';k,\omega)
-\mathrm{i}\boldsymbol{\mathcal{M}'^*}\tilde 
G_m(r,r';k,\omega)\right]\chi_{mk}^*(\theta',z'),
\label{g}
\end{eqnarray}
\end{subequations}
where we have introduced the two scalar Green's functions $F_m, G_m$, which 
satisfy
\begin{subequations}
\begin{eqnarray}
d_m\mathcal{D}_mF_m(r,r')&=&\frac{\omega^2\mu}{\varepsilon}\frac{1}{r}
\delta(r-r'),\label{difeqF}\\
d_m\mathcal{D}_mG_m(r,r')&=&\omega^2\frac{1}{r}\delta(r-r'),\label{difeqG}
\end{eqnarray}
\end{subequations}
while  $\tilde F_m$ and $\tilde G_m$ are annihilated by the operator 
$d_m\mathcal{D}_m$,
\begin{equation}
d_m\mathcal{D}_m\tilde F(r,r')=d_m\mathcal{D}_m\tilde 
G(r,r')=0.\label{homodifeq}
\end{equation}

The Green's dyadics have now the form:
\begin{subequations}
\begin{eqnarray}
\mathbf{\Gamma'(r,r'};\omega)&=&\sum_{m=-\infty}^{\infty}
\int_{-\infty}^{\infty}\frac{\d k}{2\pi}\bigg{\{}\boldsymbol{\mathcal{M 
M'^*}}\left(-\frac{d_m-k^2}{\omega^2\mu}\right)F_m(r,r')
\nonumber\\
&&\quad\mbox{}+\frac{1}{\omega}\boldsymbol{\mathcal{M 
N'^*}}\left(-\frac{d_m-k^2}{\omega^2\mu}\right)\tilde 
F_m(r,r')+\boldsymbol{\mathcal{N N'^*}}\frac{1}{\omega^2\varepsilon}G_m(r,r'){}
\nonumber\\
&&\quad\mbox{}+\frac{1}{\omega\varepsilon}\boldsymbol{\mathcal{N M'^*}}\tilde 
G_m(r,r')\bigg{\}}\chi_{mk}(\theta,z)\chi_{mk}^*(\theta',z'),
\label{gammaprimedyadic}\\
\mathbf{\Phi(r,r'};\omega)&=&\sum_{m=-\infty}^{\infty}
\int_{-\infty}^{\infty}\frac{\d k}{2\pi}\bigg{\{}
-\frac{\mathrm{i}}{\omega}\boldsymbol{\mathcal{M 
N'^*}}G_m(r,r')-\mathrm{i}\boldsymbol{\mathcal{M M'^*}}\tilde G_m(r,r'){}
\nonumber\\
&&\quad\mbox{}-\frac{\mathrm{i}\varepsilon}{\omega\mu}\boldsymbol{\mathcal{N 
M'^*}}F_m(r,r')-\frac{\mathrm{i}\varepsilon}{\omega^2\mu}\boldsymbol{\mathcal{N 
N'^*}}\tilde F_m(r,r')\bigg{\}}\nonumber\\
&&\quad\qquad\times\chi_{mk}(\theta,z)\chi_{mk}^*(\theta',z').
\label{phidyadic}
\end{eqnarray}
\end{subequations}

In the following, we will apply these equations to a dielectric-diamagnetic 
cylinder of radius $a$, where the interior of the cylinder is characterized by 
a permittivity $\varepsilon$ and permeability $\mu$, while the outside is 
vacuum, so $\varepsilon=\mu=1$ there. 
Let us consider the case that the source point is outside, $r'>a$. If the 
field point is also outside, $r,r'>a$, the scalar Green's functions $F'_m, 
G'_m, \tilde F', \tilde G'$  that make up the above Green's dyadics (we 
designate with primes the outside scalar Green's functions or 
constants) obey the differential equations (\ref{difeqF}), (\ref{difeqG}), and 
(\ref{homodifeq}) with $\varepsilon=\mu=1$.
 To solve these fourth-order differential equations we introduce auxiliary 
Green's functions $\mathcal{G}^{F'(G')}_m(r,r')$ and $\mathcal{G}^{\tilde 
F'(\tilde G')}_m(r,r')$, satisfying $(m\neq0)$
\begin{subequations}
\begin{eqnarray}
d_m\mathcal{G}^{F'(G')}_m(r,r')&=&\frac{1}{r}\delta(r-r'),\\
d_m\mathcal{G}^{\tilde F'(\tilde G')}_m(r,r')&=&0,
\end{eqnarray}
\end{subequations}
which therefore have the general form
\begin{subequations}
\begin{eqnarray}
\mathcal{G}^{F'(G')}_m(r,r')&=&a'^{F(G)}_m(r')\frac{1}{r^{|m|}}
-\frac{1}{2|m|}\left(\frac{r_<}{r_>}\right)^{|m|},\\
\mathcal{G}^{\tilde F'(\tilde G')}_m(r,r')&=&a'^{\tilde F(\tilde 
G)}_m(r')\frac{1}{r^{|m|}},
\end{eqnarray}
\end{subequations}
where $r_<(r_>)$ is the lesser (greater) of $r$, $r'$ 
and we discarded a possible 
$r^{|m|}$ term because we seek a solution which vanishes at infinity. 
Thus $F'_m, G'_m, \tilde F'$ and 
$\tilde 
G'$ satisfy the second-order differential equations
\begin{subequations}
\begin{eqnarray}
\mathcal{D}_mF'_m&=&\omega^2\mathcal{G}^{F'}_m,
\qquad\mathcal{D}_mG'_m=\omega^2\mathcal{G}^{G'}_m,\label{sec-ord-inhom}\\
\mathcal{D}_m\tilde F'_m&=&\omega^2\mathcal{G}^{\tilde F'},
\qquad\mathcal{D}_m\tilde 
G'_m=\omega^2\mathcal{G}^{\tilde G'}.\label{sec-ord-hom}
\end{eqnarray}
\end{subequations}
 Now, from 
(\ref{sec-ord-inhom}) and the first identity in (\ref{besselop}) we learn that
($\lambda^{\prime2}=\omega^2-k^2$)
\begin{equation}
F'_m-\frac{\omega^2}{\lambda'^2}\mathcal{G}^{F'}_m=A'^F_m(r')H_m(\lambda'r)
-\frac{\omega^2}{\lambda'^2}\frac{\pi}{2\mathrm{i}}J_m(\lambda'r_<)
H_m(\lambda'r_>),
\end{equation}
while $G'_m$ obeys a similar expression with the replacement  
$F\rightarrow G$. Similarly, from (\ref{sec-ord-hom})
\begin{equation}
\tilde F'_m-\frac{\omega^2}{\lambda'}\mathcal{G}^{\tilde F}_m=A'^{\tilde 
F}_m(r')H_m(\lambda'r),
\end{equation}
and for $\tilde G'_m$  replace $F\rightarrow G$. Here, to have
the appropriate outgoing-wave boundary condition at infinity, we have
used $H_m(\lambda'r)=H_m^{(1)}(\lambda'r)$.

The dependence of the constants on the second variable $r'$ can be deduced by 
noticing that, naturally, the Green's dyadics have to satisfy Maxwell's 
equations in their second variable. Thus, by imposing the Helmholtz equations
in the second variable together with the boundary conditions at 
$r'=\infty$, it is easy to see that
\begin{subequations}
\begin{eqnarray}
a'^F_m(r')&=&a'^F_m\frac{1}{r'^{|m|}}+b'^F_mH_m(\lambda'r'),\label{constanta}\\
A'^F_m(r')&=&A'^F_m\frac{1}{r'^{|m|}}+B'^F_mH_m(\lambda'r'),\label{constantA}
\end{eqnarray}
\end{subequations}
and with similar relations for $a'^G_m(r'), A'^G_m(r'), a'^{\tilde G}_m(r')$, 
and so on. Then, the outside Green's functions have the form
\begin{eqnarray}
F'_m(r,r')&=&\frac{\omega^2}{\lambda'^2}\left[\frac{a'^F_m}
{r'^{|m|}}+b'^F_mH_m(\lambda' 
r')\right]{r^{-|m|}}-\frac{\omega^2}{\lambda'^2}\frac{1}{2|m|}
\left(\frac{r_<}{r_>}\right)^{|m|}\nonumber\\
&&\mbox{}+\left[\frac{A'^F_m}{r'^{|m|}}+B'^F_mH_m
(\lambda' r')\right]H_m(\lambda' 
r)-\frac{\omega^2}{\lambda'^2}\frac{\pi}{2\mathrm{i}}J_m(\lambda' r_<)
H_m(\lambda' r_>),
\nonumber\\
\end{eqnarray}
while $G'_m$ has the same form with the constants $a'^F_m, b'^F_m, A'^F_m, 
B'^F_m$ replaced by $a'^G_m,b'^G_m,A'^G_m,B'^G_m$, respectively. The 
homogeneous differential equations have solutions
\begin{equation}
\tilde F'_m(r,r')=\frac{\omega^2}{\lambda'^2}\left[\frac{a'^{\tilde 
F}_m}{r'^{|m|}}+b'^{\tilde F}_mH_m(\lambda' 
r')\right]{r^{-|m|}}+\left[\frac{A'^{\tilde F}_m}{r'^{|m|}}+B'^{\tilde 
F}_mH_m(\lambda' r')\right]H_m(\lambda' r),
\end{equation}
while in  $\tilde G'_m$ we replace 
$a'^{\tilde F}\rightarrow a'^{\tilde G}$, etc. 

When the source point is outside and the field point is inside, all the
Green's functions satisfy the
homogeneous equations (\ref{homodifeq}) with $\varepsilon$, $\mu\ne1$, 
and then, following the above scheme we have that
\begin{equation}
d_m\mathcal{G}^F_m=d_m\mathcal{G}^G_m=d_m\mathcal{G}^{\tilde 
F}_m=d_m\mathcal{G}^{\tilde G}_m=0,
\end{equation}
and
\begin{equation}
\mathcal{G}^F(r,r')=a^F_m(r')r^{|m|},
\end{equation}
since now $r$ can be $0$. Also $\mathcal{D}_mF_m=\omega^2\mathcal{G}^F_m$ and
\begin{equation}
F_m-\frac{\omega^2}{\lambda^2}\mathcal{G}^F_m=A^F_m(r')J_m(\lambda r).
\end{equation}
$G_m, \tilde F_m$ and $\tilde G_m$ have the same form, and the constants 
$a^F_m(r'), A^F_m(r')$, etc.\ follow the pattern in (\ref{constanta}) and 
(\ref{constantA}).
Now,  we may write for $r<a,r'>a$
\begin{equation}
F_m(r,r')=\frac{\omega^2}{\lambda^2}\left[\frac{a^F_m}{r'^{|m|}}+b^F_mH_m(\lambda'r')
\right]r^{|m|}+\left[\frac{A^F_m}{r'^{|m|}}+B^F_mH_m(\lambda'r')\right]
J_m(\lambda r),
\end{equation}
and similarly for $G_m, \tilde F_m, \tilde G_m,$ with the corresponding change 
of constants. In all of the above, the outside and inside forms of 
$\lambda$ are given by
\begin{equation}
\lambda'^2=\omega^2-k^2,
\qquad
\lambda^2=\omega^2\mu\varepsilon-k^2.
\end{equation}

The various constants are to be determined, as far as possible, by the 
boundary 
conditions at $r=a$. The boundary conditions at the surface of the dielectric 
cylinder are the continuity of tangential components of the electric field, of 
the normal component of the electric displacement, of the normal component of 
the magnetic induction, and of the tangential components of the magnetic field
(we assume that there are no surface charges or currents):
\begin{eqnarray}
\mathbf{E}_t\text{ is continuous,\qquad}
&&\varepsilon E_n\text{ is continuous,}\nonumber\\
\mathbf{H}_t\text{ is continuous,\qquad}
&&\mu H_n\text{ is continuous.}\label{bc}
\end{eqnarray}
These conditions are redundant, but we will impose all of them as a check of 
consistency. In terms of the Green's dyadics, the conditions read
\begin{subequations}
\begin{eqnarray}
\boldsymbol{\hat \theta \cdot\Gamma'}\bigg |_{r=a-}^{r=a+}&=&\mathbf{0},
\label{bc1}\\
\boldsymbol{\hat z \cdot\Gamma'}\bigg |_{r=a-}^{r=a+}
&=&\mathbf{0},\label{bc2}\\
\boldsymbol{\hat r\cdot}\varepsilon\boldsymbol{\Gamma'}
\bigg |_{r=a-}^{r=a+}&=&\mathbf{0},\label{bc3}\\
\boldsymbol{\hat r\cdot}\mu\boldsymbol{\Phi}\bigg|_{r=a-}^{r=a+}&=&
\mathbf{0},\label{bc4}\\
\boldsymbol{\hat \theta\cdot\Phi}\bigg |_{r=a-}^{r=a+}&=&
\mathbf{0},\label{bc5}\\
\boldsymbol{\hat z\cdot\Phi}\bigg |_{r=a-}^{r=a+}&=&\mathbf{0}.\label{bc6}
\end{eqnarray}
\end{subequations}
We can also impose the Helmholtz equations (\ref{helmholtz1}) and 
(\ref{helmholtz2}). From those we learn that the coefficients of terms with 
powers of $r$ are related in the following way
\begin{subequations}
\begin{eqnarray}
\hat a'^F+\hat a'^G&=&0,\label{hm'1}\\
b'^G-(\mbox{sgn}m)\frac{k}{\omega}b'^{\tilde F}&=&0,\label{hm'2}\\
b'^{\tilde G}-(\mbox{sgn}m)\frac{k}{\omega}b'^ F&=&0,\label{hm'3}
\end{eqnarray}
\end{subequations}
for the Green's dyadics outside the cylinder and equivalent expressions for 
the inside (no primes)
\begin{subequations}
\begin{eqnarray}
\frac{\varepsilon}{\mu}\hat a^F-\hat a^G&=&0,\label{hm1}\\
b^G+(\mbox{sgn}m)\frac{\varepsilon}{\mu}\frac{k}{\omega}b^{\tilde 
F}&=&0,\label{hm2}\\
b^{\tilde G}+(\mbox{sgn}m)\frac{\varepsilon}{\mu}\frac{k}{\omega}b^ 
F&=&0,\label{hm3}
\end{eqnarray}
\end{subequations}
where we have introduced the abbreviations for any constant $K$
\begin{equation}
\hat K^F=K^F-(\mbox{sgn}m)\frac{k}{\omega}K^{\tilde F},\qquad\hat 
K^G=K^G-(\mbox{sgn}m)\frac{\omega}{k}K^{\tilde G}, 
\end{equation}
and the same for $\hat K'^F$ and $\hat K'^G$ (the outside).
Then, from the boundary conditions we can solve for the remaining constants. 
Notice that, due to the tensorial character of the Green's dyadics, each of 
the above six boundary conditions (\ref{bc1}), (\ref{bc2}), (\ref{bc3}), 
(\ref{bc4}), (\ref{bc5}), (\ref{bc6}) are in fact three equations 
corresponding to the three prime coordinates.
From (\ref{bc1}) we get the following three equations:
\begin{subequations}
\begin{eqnarray}
-\varepsilon\lambda aJ'_m(\lambda a)B^{\tilde 
F}_m&-&\frac{mk}{\omega\varepsilon}J_m(\lambda 
a)B^G_m=-\lambda'aH'_m(\lambda' a)B'^{\tilde F}_m\nonumber\\
&-&\frac{mk}{\omega}H_m(\lambda' 
a)B'^G_m+\frac{mk\omega}{\lambda'^2}\frac{\pi}{2\mathrm{i}}J_m(\lambda' 
a),
\label{thetagamma1}\\
-\varepsilon|m|\lambda a J'_m(\lambda 
a)A^F_m&+&\frac{mk\varepsilon}{\omega}\lambda a J'_m(\lambda a)A^{\tilde 
F}_m+\frac{m^2k^2}{\omega^2\varepsilon}J_m(\lambda 
a)A^G_m\nonumber\\
-\frac{m|m|k}{\omega\varepsilon}J_m(\lambda a)A^{\tilde G}_m&=&
-|m|\lambda'aH'_m(\lambda'a)A'^F_m+\frac{mk}{\omega}\lambda'a
H'_m(\lambda'a)A'^{\tilde 
F}_m\nonumber\\
&+&\frac{m^2k^2}{\omega^2}
H_m(\lambda'a)A'^G_m-\frac{m|m|k}{\omega}
H_m(\lambda'a)A'^{\tilde G}_m,
\label{thetagamma2}\\
\varepsilon\lambda aJ'_m(\lambda 
a)B^F_m&+&\frac{mk}{\omega\varepsilon}J_m(\lambda a)B^{\tilde 
G}_m=\lambda'aH'_m(\lambda'a)B'^F_m\nonumber\\
&-&\frac{\omega^2}{\lambda'^2}
\frac{\pi}{2\mathrm{i}}\lambda'aJ'_m(\lambda'a)
+\frac{mk}{\omega}
H_m(\lambda'a)B'^{\tilde G}_m.\label{thetagamma3}
\end{eqnarray}
\end{subequations}
The three equations following from (\ref{bc2}) are:
\begin{subequations}
\begin{eqnarray}
B^G_m&=&\varepsilon\left(\frac{\lambda'}{\lambda}\right)^2
\left[B'^G_m\frac{H_m(\lambda'a)}{J_m(\lambda 
a)}-\frac{\omega^2}{\lambda'^2}\frac{\pi}{2\mathrm{i}}
\frac{J_m(\lambda'a)}{J_m(\lambda a)}\right],\label{zgamma1}\\
|m|A^G_m-m\frac{\omega}{k}A^{\tilde 
G}_m&=&\varepsilon\left(\frac{\lambda'}{\lambda}\right)^2
\frac{H_m(\lambda'a)}{J_m(\lambda 
a)}\left[|m|A'^G_m-m\frac{\omega}{k}A'^{\tilde 
G}_m\right],\label{zgamma2}\\
B^{\tilde 
G}_m&=&\varepsilon\left(\frac{\lambda'}{\lambda}\right)^2
\frac{H_m(\lambda'a)}{J_m(\lambda 
a)}B'^{\tilde G}_m,\label{zgamma3}
\end{eqnarray}
\end{subequations}
and those coming from (\ref{bc3}) are:
\begin{subequations}
\begin{eqnarray}
m\varepsilon^2J_m(\lambda a)B^{\tilde F}_m&+&\frac{k}{\omega}\lambda a 
J'_m(\lambda a)B^G_m=mH_m(\lambda'a)B'^{\tilde 
F}_m+\frac{k}{\omega}\lambda'aH'_m(\lambda'a)B'^G_m\nonumber\\
&-&\frac{k\omega}{\lambda'^2}\frac{\pi}{2\mathrm{i}}\lambda'aJ'_m(\lambda'a),
\label{rgamma1}\\
\varepsilon^2m^2J_m(\lambda 
a)A^F_m&-&\frac{m|m|k}{\omega}\varepsilon^2J_m(\lambda 
a)A^{\tilde F}_m-\frac{k^2|m|}{\omega^2}\lambda aJ'_m(\lambda 
a)A^G_m\nonumber\\
+\frac{mk}{\omega}\lambda aJ'_m(\lambda a)A^{\tilde 
G}_m&=&m^2H_m(\lambda'a)A'^F_m-\frac{m|m|k}{\omega}
H_m(\lambda'a)A'^{\tilde F}_m\nonumber\\
&-&\frac{|m|k^2}{\omega^2}\lambda'aH'_m(\lambda'a)A'^G_m+
\frac{mk}{\omega}\lambda'aH'_m(\lambda'a)A'^{\tilde 
G}_m,\label{rgamma2}\\
\varepsilon^2m J_m(\lambda a)B^F_m&+&\frac{k}{\omega}\lambda a 
J'_m(\lambda a)B^{\tilde 
G}_m=mH_m(\lambda'a)B'^F_m-m\frac{\omega^2}{\lambda'^2}
\frac{\pi}{2\mathrm{i}}J_m(\lambda'a)\nonumber\\
&+&\frac{k}{\omega}\lambda'aH'_m(\lambda'a)B'^{\tilde G}_m.
\label{rgamma3}
\end{eqnarray}
\end{subequations}
From the set of equations involving the magnetic part, $\Phi$, we find that 
(\ref{bc4}) gives us
\begin{subequations} 
\begin{eqnarray}
\mu mJ_m(\lambda a)B^G_m&+&\frac{\varepsilon k}{\omega}\lambda a 
J'_m(\lambda a)B^{\tilde 
F}_m=mH_m(\lambda'a)B'^G_m\nonumber\\
&-&m\frac{\omega^2}{\lambda'^2}\frac{\pi}{2\mathrm{i}}J_m(\lambda'a)
+\frac{k}{\omega}\lambda'aH'_m(\lambda'a)B'^{\tilde F}_m,
\label{rphi1}\\
\frac{\mu m|m|k}{\omega}J_m(\lambda a)A^G_m&-&\mu m^2
 J_m(\lambda 
a)A^{\tilde G}_m-\frac{\varepsilon km}{\omega}\lambda aJ'_m(\lambda 
a)A^F_m\nonumber\\
+\frac{\varepsilon |m|k^2}{\omega^2}\lambda aJ'_m(\lambda a)A^{\tilde 
F}_m&=&\frac{m|m|k}{\omega}H_m(\lambda'a)A'^G_m
-m^2H_m(\lambda'a)A'^{\tilde G}_m\nonumber\\
&-&\frac{mk}{\omega}\lambda'aH'_m(\lambda'a)A'^F_m+
\frac{|m|k^2}{\omega^2}\lambda'aH'_m(\lambda'a)A'^{\tilde 
F}_m,\label{rphi2}\\
-\mu mJ_m(\lambda a)B^{\tilde G}_m&-&\frac{\epsilon k}{\omega}\lambda a
J'_m(\lambda a)B^F_m=-mH_m(\lambda'a)B'^{\tilde G}_m\nonumber\\
&+&\frac{k\omega}{\lambda'^2}\frac{\pi}{2\mathrm{i}}\lambda'a
J'_m(\lambda'a)-\frac{k}{\omega}\lambda'aH'_m(\lambda'a)B'^F_m.\label{rphi3}
\end{eqnarray}
\end{subequations}
By imposing (\ref{bc5}) we get the conditions
\begin{subequations}
\begin{eqnarray}
\lambda aJ'_m(\lambda a)B^G_m&+&
\frac{\varepsilon mk}{\omega\mu}J_m(\lambda 
a)B^{\tilde F}_m=\lambda'aH'_m(\lambda' a)B'^G_m\nonumber\\
&+&\frac{mk}{\omega}H_m(\lambda' a)B'^{\tilde 
F}_m-\frac{\omega^2}{\lambda'^2}\frac{\pi}{2\mathrm{i}}\lambda'aJ'_m(\lambda' 
a),
\label{thetaphi1}\\
-\frac{|m|k}{\omega}\lambda aJ'_m(\lambda a)A^G_m&+&m\lambda aJ'_m(\lambda 
a)A^{\tilde G}_m+\frac{m^2k\varepsilon}{\omega\mu}J_m(\lambda 
a)A^F_m\nonumber\\
-\frac{m|m|\varepsilon k^2}{\omega^2\mu}J_m(\lambda a)A^{\tilde F}_m
&=&
-\frac{|m|k}{\omega}\lambda'aH'_m(\lambda'a)A'^G_m+m\lambda'aH'_m(\lambda'a)
A'^{\tilde G}_m\nonumber\\
&+&\frac{m^2k}{\omega}H_m(\lambda'a)A'^F_m
-\frac{m|m|k^2}{\omega^2}H_m(\lambda'a)A'^{\tilde 
F}_m,
\label{thetaphi2}\\
\lambda aJ'_m(\lambda a)B^{\tilde 
G}_m&+&\frac{mk\varepsilon}{\omega\mu}J_m(\lambda 
a)B^F_m=\lambda'aH'_m(\lambda'a)B'^{\tilde G}_m\nonumber\\
&-&\frac{\omega 
mk}{\lambda'^2}\frac{\pi}{2\mathrm{i}}J_m(\lambda'a)
+\frac{mk}{\omega}H_m(\lambda'a)B'^F_m.\label{thetaphi3}
\end{eqnarray}
\end{subequations}
And finally (\ref{bc6}) gives us
\begin{subequations}
\begin{eqnarray}
B^{\tilde 
F}_m&=&\frac{\mu}{\varepsilon}\left(\frac{\lambda'}{\lambda}\right)^2
\frac{H_m(\lambda'a)}{J_m(\lambda 
a)}B'^{\tilde F}_m,\label{zphi1}\\
-A^F_m+\frac{k}{\omega}\frac{|m|}{m}A^{\tilde 
F}_m&=&\frac{\mu}{\varepsilon}\left(\frac{\lambda'}{\lambda}\right)^2
\frac{H_m(\lambda'a)}{J_m(\lambda 
a)}\left[-A'^F_m+\frac{k}{\omega}\frac{|m|}{m}A'^{\tilde 
F}_m\right],\label{zphi2}\\
B^F_m&=&\frac{\mu}{\varepsilon}\left(\frac{\lambda'}{\lambda}\right)^2
\left[B'^F_m\frac{H_m(\lambda'a)}{J_m(\lambda 
a)}-\frac{\omega^2}{\lambda'^2}\frac{\pi}{2\mathrm{i}}
\frac{J_m(\lambda'a)}{J_m(\lambda a)}\right].\label{zphi3}
\end{eqnarray}
\end{subequations}

  By combining these equations we find the remaining constants, but 
the equations are not all 
independent. First, from (\ref{zgamma2}), (\ref{zphi2}), 
(\ref{thetagamma2}) and (\ref{rgamma2}) we learn that the coefficients of 
terms involving Bessel functions and  $r'^{-|m|}$ cancel among themselves in a 
way such that the ones from the outside do not mix with those from the inside:
\begin{subequations}
\begin{eqnarray}
\hat A^F_m=\hat A^G_m&=&0,\label{bc-sol1}\\
\hat A'^F_m=\hat A'^G_m&=&0.\label{bc-sol2}
\end{eqnarray}
\end{subequations}
The same can be found if we use (\ref{thetaphi2}) and  (\ref{rphi2}) instead 
of (\ref{thetagamma2}) and (\ref{rgamma2}).

Next we determine the coefficients of functions involving just Bessel functions. 
From (\ref{thetaphi3}) and (\ref{rgamma3}) we find using 
(\ref{zphi3}) and (\ref{zgamma3}) that
\begin{subequations}
\begin{eqnarray}
B^{\tilde 
G}_m&=&-\frac{\varepsilon^2}{\mu}(1-\varepsilon\mu)
\frac{mk\omega}{\lambda\lambda'D}J_m(\lambda 
a)H_m(\lambda'a)B^F_m,\\
B'^{\tilde 
G}_m&=&-\left(\frac{\lambda}{\lambda'}\right)^2
\frac{\varepsilon}{\mu}(1-\varepsilon\mu)
\frac{mk\omega}{\lambda\lambda'D}J^2_m(\lambda 
a)B^F_m,\\
B'^F_m&=&\frac{\omega^2}{\lambda'^2}\frac{\pi}{2\mathrm{i}}\frac{J_m(\lambda'a 
)}{H_m(\lambda'a)}+\left(\frac{\lambda}{\lambda'}\right)^2
\frac{\varepsilon}{\mu}\frac{J_m(\lambda 
a)}{H_m(\lambda'a)}B_m^F,
\end{eqnarray}
\end{subequations}
all in terms of
\begin{equation}
B_m^F=-\frac{\mu}{\varepsilon}\frac{\omega^2}{\lambda\lambda'},
\frac{D}{\Xi}
\end{equation}
found by subtracting $\frac{k}{\omega}$ times equation (\ref{rgamma3}) from 
(\ref{thetaphi3}) and using (\ref{rphi3})\footnote{(\ref{rphi3}) is the same 
equation as (\ref{thetagamma3}), which can easily be seen by using 
(\ref{zphi3}).}. The denominators occurring here are\footnote{The denominator
structure appearing in $\Xi$ is precisely that given by Stratton \cite{stratton},
and is the basis for the calculation given, for example in 
Ref.~\cite{milt-nest-nest}.  It will be employed in an independent
rederivation of the Casimir energy for a dilute dielectric cylinder
\cite{romeom}.}
\begin{subequations}
\begin{eqnarray}
\Xi&=&(1-\varepsilon\mu)^2\frac{m^2k^2\omega^2}{\lambda^2\lambda^{\prime2}}
J^2_m(\lambda  a)H^2_m(\lambda'a)-D\tilde D,\\
D&=&\varepsilon\lambda'aJ'_m(\lambda a)H_m(\lambda'a)-\lambda 
aH'_m(\lambda'a)J_m(\lambda a),\\
\tilde D&=&\mu\lambda'aJ'_m(\lambda a)H_m(\lambda'a)-\lambda 
aH'_m(\lambda'a)J_m(\lambda a).
\end{eqnarray}
\end{subequations}
The second set of constants is found using (\ref{rphi1}), (\ref{thetagamma1}), 
(\ref{zphi1}) and (\ref{zgamma1}):
\begin{subequations}
\begin{eqnarray}
B^{\tilde 
F}_m&=&-\frac{\mu}{\varepsilon^2}(1-\varepsilon\mu)
\frac{mk\omega}{\lambda\lambda'\tilde 
D}J_m(\lambda a)H_m(\lambda'a)B^G_m,\\
B'^{\tilde 
F}_m&=&-\left(\frac{\lambda}{\lambda'}\right)^2
\frac{1}{\varepsilon}(1-\varepsilon\mu)\frac{mk\omega}{\lambda\lambda'\tilde 
D}J^2_m(\lambda a)B^G_m\label{bprimetilf},\\
B'^G_m&=&\frac{\omega^2}{\lambda'^2}\frac{\pi}{2\mathrm{i}}\frac{J_m(\lambda'a 
)}{H_m(\lambda'a)}+\left(\frac{\lambda}{\lambda'}\right)^2
\frac{1}{\varepsilon}\frac{J_m(\lambda 
a)}{H_m(\lambda'a)}B_m^G,
\end{eqnarray}
\end{subequations}
in terms of 
\begin{equation}
B_m^G=-\varepsilon\frac{\omega^2}{\lambda\lambda'}\frac{\tilde D}{\Xi}
\end{equation}
coming from (\ref{bprimetilf}) and (\ref{rgamma1})\footnote{By using 
(\ref{zphi1}) it can be seen that this equation is the same as 
(\ref{thetaphi1}).}.

It might be thought that $m=0$ is a special case, and indeed
\begin{equation}
\frac{1}{2|m|}\left(\frac{r_<}{r_>}\right)^{|m|}\rightarrow\frac{1}{2}
\ln\frac{r_<}{r_>},
\end{equation}
but just as the latter is correctly interpreted as the limit as 
$|m|\rightarrow0$, so the coefficients in the Green's functions turn out to be 
just the $m=0$ limits for those given above, so the $m=0$ case is properly 
incorporated.

It is now easy to check that, as a result of the conditions (\ref{hm'1}), 
(\ref{hm'2}), (\ref{hm'3}), (\ref{hm1}), (\ref{hm2}), (\ref{hm3}), 
(\ref{bc-sol1}), and (\ref{bc-sol2}), 
the terms in the Green's functions that involve powers of $r$ 
or $r'$ do not contribute to the electric or magnetic fields. As we might have 
anticipated, only the pure Bessel function terms contribute. (This observation 
was not noted in Ref.~\cite{deraad&milton1981}.)

\section{Stress on the Cylinder}
\label{sec3}
We are now in a position to calculate the pressure on the surface of the 
cylinder from the radial-radial component of the stress tensor
\begin{equation}
P=\langle T_{rr}\rangle(a-)-\langle T_{rr}\rangle(a+)
\end{equation}
where
\begin{equation}
T_{rr}=\frac{1}{2}\left[\varepsilon(E^2_{\theta}+E^2_z-E^2_r)
+\mu(H^2_{\theta}+H^2_z-H^2_r)\right].\label{r-rstresscomp}
\end{equation}
As a result of the boundary conditions (\ref{bc}), the pressure on the 
cylindrical walls are given by the expectation value of the squares of field 
components just outside the cylinder, therefore
\begin{eqnarray}
T_{rr}\big|_{r=a-}-T^{rr}\big|_{r=a+}&=&\frac{\varepsilon-1}{2}
\left(E^2_{\theta}+E^2_z+\frac{E^2_r}{\varepsilon}\right)\bigg|_{r=a+}
\nonumber\\
&&\quad\mbox{}+\frac{\mu-1}{2}\left(H^2_{\theta}+H^2_z
+\frac{H^2_r}{\mu}\right)\bigg|_{r=a+}.\label{pressure}
\end{eqnarray}
These expectation values are given by (\ref{vevE}), (\ref{vevH}), where the 
latter may also be written as
\begin{equation}
\langle\mathbf{H(r)\,H(r')}\rangle=-\frac{\hbar}{\omega\mu}\mathbf{\Phi(r,r')}
\times\boldsymbol{\overleftarrow\nabla'}.\label{vevH2}
\end{equation}

It is quite straightforward to write the vacuum expectation values of the 
fields occurring here outside the cylinder in terms of the Green's functions,
\begin{subequations}
\begin{eqnarray}
\langle E_r(r)E_r(r')\rangle&=&\frac{\hbar}{\mathrm{i}}\Gamma_{rr'}
=\frac{\hbar}{2\pi 
\mathrm{i}}\sum_{m=-\infty}^{\infty}\int_{-\infty}^{\infty}\frac{\d k}{2\pi}
\bigg\{-\frac{m^2}{rr'}\frac{d_m-k^2}{\omega^2}F'_m(r,r')\nonumber\\
&&\quad\mbox{}-\frac{mk}{\omega r}\frac{\partial}{\partial r'}
\frac{d_m-k^2}{\omega^2}\tilde 
F'_m(r,r')+\frac{k^2}{\omega^2}\frac{\partial}{\partial 
r}\frac{\partial}{\partial r'}G'_m(r,r)\nonumber\\
&&\quad\mbox{}+\frac{km}{\omega 
r'}\frac{\partial}{\partial r}\tilde G'_m(r,r')\bigg{\}},
\label{vev-er}\\
\langle 
E_{\theta}(r)E_{\theta}(r')\rangle&=&\frac{\hbar}{\mathrm{i}}
\Gamma_{\theta\theta'}=
\frac{\hbar}{2\pi 
\mathrm{i}}\sum_{m=-\infty}^{\infty}\int_{-\infty}^{\infty}\frac{\d k}{2\pi}
\bigg\{-\frac{\partial}{\partial 
r}\frac{\partial}{\partial r'}\frac{d_m-k^2}{\omega^2}F'_m(r,r')\nonumber\\
&&\quad\mbox{}-\frac{mk}{\omega r'}\frac{\partial}{\partial r}
\frac{d_m-k^2}{\omega^2}\tilde 
F'_m(r,r')+\frac{m^2k^2}{\omega^2 rr'}G'_m(r,r')\nonumber\\
&&\quad\mbox{}+\frac{mk}{\omega 
r}\frac{\partial}{\partial r'}\tilde G'_m(r,r')\bigg{\}},
\label{vev-etheta}\\
\langle E_z(r)E_z(r')\rangle&=&\frac{\hbar}{\mathrm{i}}
\Gamma_{zz'}=\frac{\hbar}{2\pi 
\mathrm{i}}\sum_{m=-\infty}^{\infty}\int_{-\infty}^{\infty}\frac{\d k}{2\pi}
\frac{1}{\omega^2}d_md'_mG'_m(r,r').\label{vev-ez}
\end{eqnarray}
\end{subequations}
According to (\ref{vevH2}) the magnetic field expectation
values can be written as follows,
\begin{subequations}
\begin{eqnarray}
\langle 
H_r(r)H_r(r')\rangle&=&-\frac{\hbar}{2\pi}\frac{\mathrm{i}}{\omega}
\sum_{m=-\infty}^{\infty}\int_{-\infty}^{\infty}\frac{\d k}{2\pi}
\bigg\{-\frac{m^2}{\omega 
rr'}(d'_m-k^2)G'_m(r,r')\nonumber\\
&&\quad\mbox{}+\frac{mk}{r}\frac{\partial}{\partial r'}\tilde 
G'_m(r,r')+\frac{k^2}{\omega}\frac{\partial}{\partial 
r}\frac{\partial}{\partial r'}F'_m(r,r)\nonumber\\
&&\quad\mbox{}-\frac{km}{\omega^2 
r'}\frac{\partial}{\partial r}(d'_m-k^2)\tilde F'_m(r,r')\bigg{\}},
\label{vev-hr}\\
\langle 
H_{\theta}(r)H_{\theta}(r')\rangle&=&-\frac{\hbar}{2\pi}\frac{\mathrm{i}}
{\omega}
\sum_{m=-\infty}^{\infty}\int_{-\infty}^{\infty}\frac{\d k}{2\pi}
\bigg\{-\frac{1}{\omega}\frac{\partial}{\partial 
r}\frac{\partial}{\partial r'}(d'_m-k^2)G'_m(r,r')\nonumber\\
&&\quad\mbox{}+\frac{mk}{r'}\frac{\partial}{\partial r}\tilde 
G'_m(r,r')+\frac{m^2k^2}{\omega rr'}F'_m(r,r')\nonumber\\
&&\quad\mbox{}-\frac{mk}{\omega^2 
r}\frac{\partial}{\partial r'}(d_m-k^2)\tilde F'_m(r,r')\bigg{\}},
\label{vev-htheta}\\
\langle 
H_z(r)H_z(r')\rangle&=&-\frac{\hbar}{2\pi}\frac{\mathrm{i}}{\omega}
\sum_{m=-\infty}^{\infty}\int_{-\infty}^{\infty}\frac{\d k}{2\pi}
\frac{1}{\omega}d_md'_mF'_m(r,r').
\label{vev-hz}
\end{eqnarray}
\end{subequations}
When these vacuum expectation values are substituted into the stress 
expression (\ref{pressure}),  and the property of $d_m$ exploited,
\begin{equation}
d_mr^{\pm|m|}=0,\qquad d_mJ_m(\lambda r)=-\lambda^2J_m(\lambda r),
\end{equation}
(of course, the later formula holds for $H_m$ as well and the same for $d'_m$ 
acting on the primed coordinate),
we obtain the pressure on the cylinder as
\begin{eqnarray}
P&=&
\hbar\frac{\varepsilon-1}{4\pi \mathrm{i}}
\sum_{m=-\infty}^{\infty}\int_{-\infty}^\infty\frac{\d \omega}{2\pi}
\int_{-\infty}^{\infty}
\frac{\d k}{2\pi}\frac{\lambda^2}{\Xi}\bigg{\{}H'^2_m(x')J_m(x)J'_m(x)
\lambda\lambda'x'(\omega^2\mu+k^2)\nonumber\\
&&\quad\mbox{}+H'_m(x')J^2_m(x)H_m(x')\bigg[\frac{m^2k^2\omega^2}
{x'\varepsilon}
\bigg((2\varepsilon+2)(1-\varepsilon\mu)\nonumber\\
&&\qquad\mbox{}+\frac{\omega^2\varepsilon+k^2}
{\lambda^2}(1-\varepsilon\mu)^2\bigg)
+x\lambda\lambda'\left(\frac{m^2}{x'^2}\left(k^2+
\frac{\omega^2}{\varepsilon}\right)+\lambda'^2\right)\bigg]\nonumber\\
&&\quad\mbox{}-H'_m(x')J'^2_m(x)H_m(x')\mu\lambda'^2x'
(\omega^2\varepsilon+k^2)\nonumber\\
&&\quad\mbox{}-J_m(x)J'_m(x)H^2_m(x')\lambda\lambda'x'\left[
\frac{m^2}{x'^2}(k^2\mu+\omega^2)+\lambda'^2\mu\right]\bigg{\}}\nonumber\\
&&\mbox{}+\hbar\frac{\mu-1}{4\pi\mathrm{i}}
\sum_{m=-\infty}^{\infty}\int_{-\infty}^{\infty}
\frac{\d k}{2\pi}
\frac{\lambda^2}{\Xi}\bigg{\{}(\varepsilon\longleftrightarrow\mu)\bigg{\}},
\label{euclpressure}
\end{eqnarray}
where $x=\lambda a$, $x'=\lambda'a$ and the last bracket indicates that the 
expression there is similar to the one for the electric part by switching 
$\varepsilon$ and $\mu$, showing manifest symmetry between the electric and 
magnetic parts. 

In order to simplify this expression, we make an Euclidean 
rotation \cite{euclidean},
\begin{equation}
\omega\rightarrow \mathrm{i}\zeta\qquad\lambda\rightarrow \mathrm{i}\kappa,
\end{equation}
so that the Bessel functions are replaced by the modified Bessel functions,
\begin{equation}
J_m(x)H_m(x')\rightarrow\frac{2}{\pi \mathrm{i}}I_m(y)K_m(y'),
\end{equation}
where $y=\kappa a$ and $y'=\kappa'a$. Then (\ref{euclpressure}) becomes
\begin{eqnarray}
P&=&\frac{\varepsilon-1}{16\pi^3a^4}\sum_{m=-\infty}^{\infty}
\int_{-\infty}^{\infty}\d\zeta 
a\,\d ka\frac{\hbar}{\tilde\Xi}\bigg{\{}K'^2_m(y')I_m(y)I'_m(y)y(k^2a^2
-\zeta^2a^2\mu)\nonumber\\
&&\quad\mbox{}-K'_m(y')I^2_m(y)K_m(y')\bigg[\frac{m^2k^2a^2\zeta^2a^2}
{y'^3\varepsilon}\bigg(-2(\varepsilon+1)(1-\varepsilon\mu)\nonumber\\
&&\quad\mbox{}+\frac{k^2a^2-\zeta^2a^2\varepsilon}{y^2}
(1-\varepsilon\mu)^2\bigg)-\frac{y^2}{y'}\left(\frac{m^2}{y'^2}\left(k^2a^2
-\frac{\zeta^2a^2}{\varepsilon}\right)+y'^2\right)\bigg]\nonumber\\
&&\quad\mbox{}-K'_m(y')I'^2_m(y)K_m(y')\mu y'(k^2a^2-\zeta^2a^2\varepsilon)
\nonumber\\
&&\quad\mbox{}-I_m(y)I'_m(y)K^2_m(y')y\left[\frac{m^2}{y'^2}(k^2a^2\mu
-\zeta^2a^2)+y'^2\mu\right]\bigg{\}}+(\varepsilon\leftrightarrow\mu),\nonumber\\
\label{rotpressure}
\end{eqnarray} 
where
\begin{subequations}
\begin{eqnarray}
\tilde\Xi&=&\frac{m^2k^2a^2\zeta^2a^2}{y^2y'^2}I^2_m(y)K^2_m(y')
(1-\varepsilon\mu)^2+\Delta\tilde\Delta,\label{fulldenom}\\
\Delta&=&\varepsilon y'I'_m(y)K_m(y')-yK'_m(y')I_m(y)\\
\tilde\Delta&=&\mu y'I'_m(y)K_m(y')-yK'_m(y')I_m(y)
\end{eqnarray}
\end{subequations}

This result reduces to the well-known expression for the Casimir pressure when 
the speed of light is the same inside and outside the cylinder, that is, when 
$\varepsilon\mu=1$. Then, it  is easy to see that the denominator reduces to 
\begin{equation}
\tilde\Xi=\Delta\tilde\Delta=\frac{(\varepsilon+1)^2}{4\varepsilon}
\left[1-\xi^2y^2[(I_mK_m)']^2\right],
\end{equation}
where $\xi=(\varepsilon-1)/(\varepsilon+1)$. In the numerator introduce polar 
coordinates,
\begin{equation}
y^2=k^2a^2+\zeta^2a^2,\qquad ka=y\sin\theta,\qquad\zeta 
a=y\cos\theta,\label{polarcoord}
\end{equation}
and carry out the trivial integral over $\theta$. The result is
\begin{equation}
P=-\frac{1}{8\pi^2a^4}\int_{0}^{\infty}\d y\,y^2\sum_{m=-\infty}^{\infty}
\frac{\d}{\d y}\mbox{ln}\left(1-\xi^2[y(K_mI_m)']^2\right),
\label{dildieldimag}
\end{equation}
which is exactly  the finite result derived in Ref.~\cite{milt-nest-nest}, and 
analyzed in a number of papers \cite{brevik-nyland,Gosdz-romeo,klich-romeo}.

\section{Bulk Casimir Stress}
\label{sec4}
The expression derived above, (\ref{rotpressure}), is incomplete. It contains an 
unobservable ``bulk'' energy contribution, which the formalism would give if 
either medium, that of the interior with dielectric constant $\varepsilon$ and 
permeability $\mu$, or that of the exterior with dielectric constant and 
permeability unity, fills all the space 
\cite{miltonng97}. The corresponding stresses are 
computed from the free Green's functions which satisfy (\ref{difeqF}) and 
(\ref{difeqG}), therefore 
\begin{equation}
F^{(0)}_m(r,r')=\frac{\mu}{\varepsilon}G^{(0)}_m(r,r')=
-\frac{\omega^2\mu}{\varepsilon\lambda^2}\left[\frac{1}{2|m|}
\left(\frac{r_<}{r_>}\right)^{|m|}+\frac{\pi}{2\mathrm{i}}J_m(\lambda 
r_<)H_m(\lambda r_>)\right],
\label{freegreensfunct}
\end{equation}
where $0<r,r'<\infty$. Notice that in this case, both $\tilde F^{(0)}_m$ and 
$\tilde G^{(0)}_m$ are zero and the Green's dyadics are given by
\begin{subequations}
\begin{eqnarray}
\mathbf{\Gamma}^{(0)\prime}(\mathbf{r,r'};\omega)&=&\sum_{m=-\infty}^{\infty}
\int_{-\infty}^{\infty}\frac{\d k}{2\pi}\bigg{\{}\boldsymbol{\mathcal{M 
M'^*}}\left(-\frac{d_m-k^2}{\omega^2\mu}\right)F^{(0)}_m(r,r')
\nonumber\\
&&\qquad\mbox{}+\frac{1}{\omega^2\varepsilon}\boldsymbol{\mathcal{N 
N'^*}}G^{(0)}_m(r,r')\bigg{\}}
\chi_{mk}(\theta,z)\chi_{mk}^*(\theta',z'),
\label{freegammaprimedyadic}\\
\mathbf{\Phi}^{(0)}(\mathbf{r,r'};\omega)&=&\sum_{m=-\infty}^{\infty}
\int_{-\infty}^{\infty}\frac{\d k}{2\pi}\bigg{\{}-\frac{\mathrm{i}}{\omega}
\boldsymbol{\mathcal{M 
N'^*}}G^{(0)}_m(r,r')\nonumber\\
&&\qquad\mbox{}-\frac{\mathrm{i}\varepsilon}{\omega\mu}\boldsymbol{\mathcal{N 
M'^*}}F^{(0)}_m(r,r')\bigg{\}}\chi_{mk}(\theta,z)\chi_{mk}^*(\theta',z').
\label{freephidyadic}
\end{eqnarray}
\end{subequations}
It should be noticed that such Green's dyadics do not satisfy the appropriate 
boundary conditions, and therefore we cannot use (\ref{pressure}), but rather 
one must compute the interior and exterior stresses individually by using 
(\ref{r-rstresscomp}). Because the two scalar Green's functions  differ only 
by a factor of $\mu/\varepsilon$ in this case, for the electric part the 
inside stress tensor is
\begin{eqnarray}
T^{(0)}_{rr}(a-)&=&\frac{\hbar}{2\pi 
\mathrm{i}}\sum_{m=-\infty}^{\infty}\int_{-\infty}^{\infty}
\frac{\d\omega}{2\pi}
\int_{-\infty}^{\infty}\frac{\d k}{2\pi}\frac{1}{\omega^2\varepsilon}
\bigg[\frac{\partial}{\partial 
r}\frac{\partial}{\partial r'}(-d_mG^{(0)}_m)\nonumber\\
&&\quad\mbox{}+\left(-d'_m-\frac{m^2}{rr'}\right)(-d_mG^{(0)}_m)\bigg]
\bigg{|}_{r=r'=a-},
\label{insidefreestress}
\end{eqnarray}
while the outside bulk stress is given by the same expression with 
$\lambda\rightarrow\lambda'=\omega^2-k^2$ and $\varepsilon=\mu=1$. When we 
substitute the appropriate interior and exterior Green's functions given in 
(\ref{freegreensfunct}), and  perform the Euclidean rotation, 
$\omega\rightarrow i\zeta$, we find a rather simple formula for the bulk 
contribution to the pressure
\begin{eqnarray}
P^b&=&T^{(0)}_{rr}(a-)-T^{(0)}_{rr}(a+)\nonumber\\
&=&\frac{\hbar}{16\pi^3a^4}\sum_{m=-\infty}^{\infty}\int_{-\infty}^{\infty}
\d\zeta a\, \d k a\bigg{\{}y^2I'_m(y)K'_m(y)-(y^2+m^2)I_m(y)K_m(y)\nonumber\\
&&\quad\mbox{}-y'^2I'_m(y')K'_m(y')+(y'^2+m^2)I_m(y')K_m(y')
\bigg{\}}.
\label{bpressure}
\end{eqnarray}
This term must be subtracted from the pressure given in (\ref{rotpressure}). 
Note that this term is the direct analog of that found in the case of a 
dielectric sphere in Ref.~\cite{milton1980}. Note also that $P^b=0$ in the 
special case $\varepsilon\mu=1$.

In the following, we are going to be interested in dilute dielectric media, 
where $\mu=1$ and $|\varepsilon-1|\ll1$. 
We easily find that when the integrand 
in (\ref{bpressure}) is expanded in powers of $(\varepsilon-1)$ the leading 
terms yield
\begin{eqnarray}
P^b&=&-\frac{\hbar}{8\pi^3a^4}\sum_{m=-\infty}^{\infty}
\int_{-\infty}^{\infty}\d ka\int_{-\infty}^{\infty}\d\zeta 
a\bigg{\{}(\varepsilon-1)\zeta^2a^2I_m(y)K_m(y)\nonumber\\
&&\quad\mbox{}+\frac14(\varepsilon-1)^2\frac{(\zeta a)^4}y
[I_m(y)K_m(y)]'+\mbox{O}\big((\varepsilon-1)^3
\big)\bigg]\nonumber\\
&=&-\frac{\hbar}{8\pi^2a^4}\sum_{m=-\infty}^{\infty}\int_{0}^{\infty}\d y 
\,y^3\bigg[(\varepsilon-1)I_m(y)K_m(y)\nonumber\\
&&\quad\mbox{}+\frac{3(\varepsilon-1)^2}{16}y[I_m(y)K_m(y)]'
+\mbox{O}\big((\varepsilon-1)^3\big)\bigg],\label{bulkexpandpressure}
\end{eqnarray}
where in the last form we have introduced polar coordinates as in 
(\ref{polarcoord}) and performed the angular integral.

\section{Dilute Dielectric Cylinder}
\label{sec5}
We now turn to the case of a dilute dielectric medium filling the cylinder, 
that is, set $\mu=1$ and consider $\varepsilon-1$ as small. We can then expand 
the integrand in (\ref{rotpressure}) in powers of $(\varepsilon-1)$ and, 
because the expression is already proportional to that factor, we need only 
expand the integrand to first order. Let us write it as
\begin{equation}
P\approx\frac{(\varepsilon-1)\hbar}{16\pi^3a^4}\int_{-\infty}^{\infty}\d\zeta 
a\int_{-\infty}^{\infty}\d ka\sum_{m=-\infty}^{\infty}
\frac{N}{\Delta\tilde\Delta},
\end{equation}
where we have noted that the  $(\varepsilon-1)^2$ in $\tilde\Xi$ 
(\ref{fulldenom}) can be dropped. Expanding the numerator and denominator 
according to
\begin{equation}
N=N^{(0)}+(\varepsilon-1)N^{(1)}+\dots,\qquad\Delta\tilde\Delta=1
+(\varepsilon-1)\Delta^{(1)}+\dots,
\end{equation}
we can write
\begin{equation}
P\approx\frac{(\varepsilon-1)\hbar}{16\pi^3a^4}\int_{-\infty}^{\infty}\d\zeta 
a\int_{-\infty}^{\infty}\d ka\sum_{m=-\infty}^{\infty}\bigg{\{}N^{(0)}+
(\varepsilon-1)\big(N^{(1)}-N^{(0)}\Delta^{(1)}\big)+\dots\bigg{\}},
\label{generalpexpans}
\end{equation}
where
\begin{subequations}
\begin{eqnarray}
N^{(0)}&=&-(k^2a^2-\zeta^2a^2)K'_m(y)I'_m(y)\nonumber\\
&&\quad\mbox{}-\left[\frac{m^2}{y^2}
(k^2a^2-\zeta^2a^2)+y^2\right]K_m(y)I_m(y),
\label{no}\\
N^{(1)}&=&\frac{\zeta^2a^2}{2}\left(1+\frac{m^2}{y^2}\right)
(k^2a^2-\zeta^2a^2)K'^2_m(y)I^2_m(y)\nonumber\\
&&\quad\mbox{}+\frac{\zeta^2a^2}{2}(k^2a^2-\zeta^2a^2)K'^2_m(y)I'^2_m(y)
\nonumber\\
&&\quad\mbox{}-\frac{\zeta^2a^2}{2}\left[\frac{m^2}{y^2}(k^2a^2-\zeta^2a^2)
+y^2\right]K^2_m(y)I'^2_m(y)\nonumber\\
&&\quad\mbox{}-\frac{\zeta^2a^2}{2}\left(1+\frac{m^2}{y^2}\right)
\left[\frac{m^2}{y^2}(k^2a^2-\zeta^2a^2)+y^2\right]K^2_m(y)I^2_m(y)
\nonumber\\
&&\quad\mbox{}+\zeta^2a^2\left[y\left(1+\frac{m^2}{y^2}\right)
+\frac{m^2}{y^3}(k^2a^2-\zeta^2a^2)-\frac{4}{y^3}m^2k^2a^2\right]
\nonumber\\
&&\qquad\times
K'_m(y)K_m(y)I^2_m(y)\nonumber\\
&&\quad\mbox{}+\left[y^2\zeta^2a^2-\zeta^2a^2(k^2a^2-\zeta^2a^2)\right]
K_m(y)K'_m(y)I_m(y)I'_m(y)\nonumber\\
&&\quad\mbox{}+\left[y\zeta^2a^2+\frac{\zeta^2a^2}{y}(k^2a^2-\zeta^2a^2)\right]
K_m(y)K'_m(y)I'^2_m(y),
\label{n1}\\
\Delta^{(1)}&=&-\frac{1}{y}\zeta^2a^2[I_m(y)K_m(y)]'+yI'_m(y)K_m(y)-
\zeta^2a^2I'_m(y)K'_m(y)\nonumber\\
&&\quad\mbox{}+\zeta^2a^2\left(1+\frac{m^2}{y^2}\right)I_m(y)K_m(y).
\label{delta1}
\end{eqnarray}
\end{subequations}
When we introduce polar coordinates as in (\ref{polarcoord})  and perform the 
trivial angular integrals,
the straightforward reduction of (\ref{generalpexpans}) is
\begin{eqnarray}
P&\approx&-\frac{\hbar}{8\pi^2a^4}(\varepsilon-1)\sum_{m=-\infty}^{\infty}
\int_{0}^{\infty}\d y\bigg{\{}y^3K_m(y)I_m(y)\nonumber\\
&&\quad\mbox{}-(\varepsilon-1)\frac{y^4}{2}\bigg[\frac{1}{2}
K'^2_m(y)I'_m(y)I_m(y)\nonumber\\
&&\quad\mbox{}+K'_m(y)I'^2_m(y)K_m(y)+K'^2_m(y)I'^2_m(y)\frac{y}{4}
\nonumber\\
&&\quad\mbox{}-K'^2_m(y)I^2_m(y)\frac{y}{4}\left(1+\frac{m^2}{y^2}\right)
+K^2_m(y)I^2_m(y)\frac{y}{2}\left(1+\frac{m^2}{y^2}\right)
\left(1-\frac{m^2}{2y^2}\right)\nonumber\\
&&\quad\mbox{}+K^2_m(y)I'_m(y)I_m(y)\left(1+\frac{m^2}{2y^2}\right)
-K^2_m(y)I'^2_m(y)\frac{y}{2}\left(1-\frac{m^2}{2y^2}\right)\bigg]\bigg{\}}.
\nonumber\\
\label{p1st&2ndorder}
\end{eqnarray}
The leading term in the pressure,
\begin{equation}
P^{(1)}=-\frac{\hbar}{8\pi^2a^4}(\varepsilon-1)
\sum_{m=-\infty}^{\infty}\int_{0}^{\infty}\d y\,y^3K_m(y)I_m(y),
\end{equation}
can also be obtained from (\ref{rotpressure}) by setting $\varepsilon=\mu=1$ 
everywhere in the integrand, and the denominator $\tilde\Xi$ is then unity. 
This is also exactly what is obtained to leading order O$[(\varepsilon-1)^1]$ 
from the bulk stress (\ref{bulkexpandpressure}).  Thus the total stress
vanishes in leading order:
\begin{equation}
P^{(1)}-P^b=\mbox{O}[(\varepsilon-1)^2],
\end{equation}
which is consistent with the interpretation of the Casimir energy as arising 
from the pairwise interaction of dilutely distributed molecules.

\section{Evaluation of the $(\varepsilon-1)^2$ term}
\label{sec6}
We now turn to the considerably more complex evaluation of the 
$(\varepsilon-1)^2$ term in (\ref{p1st&2ndorder}).
\subsection{Summation method}
As a first approach to evaluating this second-order term,
we first carry out the sum on $m$ by use of the addition theorem
\begin{equation}
K_0(kP)=\sum_{m=-\infty}^\infty \e^{\mathrm{i}m(\phi-\phi')}K_m(k\rho)I_m(k\rho'),
\quad \rho>\rho',
\label{addthm}
\end{equation}
where $P=\sqrt{\rho^2+\rho^{\prime2}-2\rho\rho'\cos(\phi-\phi')}$.
Then by squaring this addition theorem and applying suitable differential
operators, in the singular limit $\rho'\to\rho$ we obtain the following
formal results:
\begin{subequations}
\begin{eqnarray}
\sum_{m=-\infty}^\infty K^2_m(k\rho)I^2_m(k\rho)&=&
\int_0^{2\pi}\frac{\d\phi}{2\pi}K^2_0(z),\\
\sum_{m=-\infty}^\infty m^2K^2_m(k\rho)I^2_m(k\rho)&=&
\int_0^{2\pi}\frac{\d\phi}{2\pi}[K'_0(z)]^2(k\rho)^2\cos^2\frac\phi2,
\\
\sum_{m=-\infty}^\infty m^4K^2_m(k\rho)I^2_m(k\rho)&=&
\int_0^{2\pi}\frac{\d\phi}{2\pi}
\left[K'_0(z)\frac{z}4
-K''_0(z)(k\rho)^2\cos^2\frac\phi2\right]^2,\nonumber\\
\\
\sum_{m=-\infty}^\infty K^2_m(k\rho)I_m(k\rho)I_m'(k\rho)&=&
\int_0^{2\pi}\frac{\d\phi}{2\pi}K_0(z)
K_0'(z)\sin\frac\phi2,\\
\sum_{m=-\infty}^\infty m^2K^2_m(k\rho)I_m(k\rho)I_m'(k\rho)&=&
\int_0^{2\pi}\frac{\d\phi}{2\pi}\frac{z}2 K_0(z)K'_0(z)k\rho\cos^2\frac\phi2,
\\
\sum_{m=-\infty}^\infty K^{\prime2}_m(k\rho)I^2_m(k\rho)&=&
\int_0^{2\pi}\frac{\d\phi}{2\pi}
\left[K'_0(z)\right]^2\sin^2\frac\phi2,\\
\sum_{m=-\infty}^\infty m^2 K_m^2(k\rho)I_m^{\prime2}(k\rho)&=&
\int_0^{2\pi}\frac{\d\phi}{2\pi}\frac{z^2}4K^2_0(z)\cos^2\frac\phi2,
\\
\sum_{m=-\infty}^\infty I_m(k\rho)I_m'(k\rho)K_m^{\prime2}(k\rho)
&=&\int_0^{2\pi}\frac{\d\phi}{2\pi}
K_0'(z)\sin\frac\phi2\nonumber\\
&&\quad\times\left[K_0(z)
\sin^2\frac\phi2-\frac{K'_0(z)}{z}\right],\nonumber\\
\\
\sum_{m=-\infty}^\infty I_m^{\prime2}(k\rho)K_m^{\prime2}(k\rho)&=&
\int_0^{2\pi}\frac{\d\phi}{2\pi}
\left[K_0(z)\sin^2\frac\phi2-\frac{K_0'(z)}{z}\right]^2.\nonumber\\
\end{eqnarray}
\end{subequations}
Here $z=2k\rho\sin\frac\phi2$, and we recognize that in this
singular limit (which omits delta functions, i.e., contact terms) terms
with $I_m$ and $K_m$ interchanged in the sum have the same values. (For
further discussion of this, see Ref.~\cite{romeom}.)

When we put this all together, we obtain the following expression
for the pressure at second order:
\begin{eqnarray}
P^{(2)}&=&\frac{(\varepsilon-1)^2}{4096\pi^2a^4}\int_0^\infty
\d z\,z^5\int_0^{2\pi}
\frac{\d\phi}{2\pi}\bigg\{\frac{K^{\prime2}_0(z)+K_0^{2}(z)
(1-4/z^2)}{\sin^6\phi/2}\nonumber\\
&&\quad\mbox{}+2\frac{(1-8/z^2)K_0^2(z)-
2(1+3/z^2)K_0^{\prime2}
(z)}{\sin^4\phi/2}-16\frac{K^2_0(z)}{z^4\sin^2\phi/2}\bigg\}\nonumber\\
&=&\frac{(\varepsilon-1)^2}{15360\pi^2a^4}\int_0^{2\pi}\frac{\d\phi}{2\pi}
\left[\frac5{\sin^6\phi/2}-\frac{66}{\sin^4\phi/2}-\frac{20}{\sin^2\phi/2}
\right].
\label{diluteresult}
\end{eqnarray}
Of course, the $\phi$ integrals in (\ref{diluteresult}) are divergent.
However, we will regulate them by continuing from 
the region where the integrals converge:
\begin{equation}
\int_0^{2\pi}\d\phi\left(\sin\frac\phi2\right)^s=\frac{2\sqrt{\pi}\Gamma\left(
\frac{1+s}2\right)}{\Gamma\left(1+\frac{s}2\right)},
\label{ancont}
\end{equation}
which is valid for $\mbox{Re}\,s>-1$.  We will take the right side of 
(\ref{ancont}) to define the angular integral for negative $s$.  Then
we see that those integrals vanish when $s=-2n$ where $n$ is a positive
integer.  Thus, this analytic continuation procedure says that the
result (\ref{diluteresult}) is zero. As for the bulk term, the addition
theorem (\ref{addthm}) implies that the $y$ integral in the second term in
 (\ref{bulkexpandpressure}) reduces to
\begin{equation}
\sum_{m=-\infty}^\infty \int_0^\infty \d y\,y^4\left(I_m(y)K_m(y)\right)'
=\int_0^\infty \d y\,y^4\frac{\d}{\d y}K_0(0)=0.
\label{bulkzero}
\end{equation}

 This argument is exactly that
given in Ref.~\cite{casmono} to show that the Casimir energy of a
dilute dielectric-diamagnetic cylinder with $\varepsilon\mu=1$ vanishes.
However, it is not very convincing, because it seems to show no relevance of
cancellations between various terms in the expressions for the pressure.
That relevance will be established in the method which follows.

\subsection{Numerical analysis}
We now turn to a detailed numerical treatment of the second-order terms in
(\ref{p1st&2ndorder}) 
and (\ref{bulkexpandpressure}).  It is based on use
of the uniform asymptotic or Debye expansions for the Bessel functions,
$m\gg1$:
\begin{subequations}
\begin{eqnarray}
I_m(y)&\sim&\frac1{\sqrt{2\pi m}}t^{1/2}\e^{m\eta}\left(1+\sum_{k=1}^\infty
\frac{u_k(t)}{m^k}\right),\\
K_m(y)&\sim&\sqrt{\frac{\pi}{2 m}}t^{1/2}\e^{-m\eta}
\left(1+\sum_{k=1}^\infty(-1)^k\frac{u_k(t)}{m^k}\right),\\
I'_m(y)&\sim&\frac1{\sqrt{2\pi m}}\frac1z t^{-1/2}\e^{m\eta}
\left(1+\sum_{k=1}^\infty\frac{v_k(t)}{m^k}\right),\\
K'_m(y)&\sim&-\sqrt{\frac{\pi}{2 m}}\frac1zt^{-1/2}\e^{-m\eta}
\left(1+\sum_{k=1}^\infty(-1)^k\frac{v_k(t)}{m^k}\right),
\end{eqnarray}
\end{subequations}
where $y=mz$ and $t=1/\sqrt{1+z^2}$. (The value of $\eta$ is irrelevant here.)
The polynomials in $t$ appearing here are generated by
\begin{subequations}
\begin{eqnarray}
u_0(t)&=&1,\quad v_0(t)=1,\\
u_k(t)&=&\frac12t^2(1-t^2)u'_{k-1}(t)+\frac18\int_0^t\d x\,(1-5x^2)u_{k-1}(x),\\
v_k(t)&=&u_k(t)+t(t^2-1)\left(\frac12 u_{k-1}(t)+t u'_{k-1}(t)\right).
\end{eqnarray}
\end{subequations}

Now suppose we write the second-order expression for the pressure as
\begin{equation}
P=\frac{(\varepsilon-1)^2}{16\pi^2a^4}\sum_{m=0}^\infty{}'\int_0^\infty \d y\,
y^4 g_m(y),
\label{epsilon2}
\end{equation}
where the explicit form  for $g_m(y)$ can be immediately read off from
(\ref{p1st&2ndorder}) and (\ref{bulkexpandpressure}), 
and the prime on the summation sign means that the 
$m=0$ term is counted with half weight.  We have recognized that the
summand is even in $m$.  Let us subtract and add the first
five terms in the uniform asymptotic expansion for $g_m$, $m\gg1$:
\begin{equation}
g_m(y)\sim \frac1{2m^2}\sum_{k=1}^5\frac1{m^k}f_k(z),
\end{equation}
where $z=y/m$ and
\begin{subequations}
\begin{eqnarray}
f_1(z)&=&\frac{4+z^2}{4z(1+z^2)^3},\\
f_2(z)&=&\frac{-8+8z^2+z^4}{8z(1+z^2)^{7/2}},\\
f_3(z)&=&\frac{16-84z^2+84z^4-16z^6-5z^8}{16z(1+z^2)^{6}},\\
f_4(z)&=&\frac{-64+1024z^2-1864z^4+504z^6-9z^8}{64z(1+z^2)^{13/2}},\\
f_5(z)&=&\frac{64-2416z^2+11808z^4-15696z^6+6856z^8-555z^{10}-15z^{12}}
{64z(1+z^2)^9}.\nonumber\\
\end{eqnarray}
\end{subequations}
We note that when these functions are inserted into (\ref{epsilon2}) in
place of $g_m$, the first three $f_k$ give divergent integrals, logarithmically
so for $f_1$ and $f_3$, and linearly divergent for $f_2$.  We also note
the crucial fact that 
\begin{equation}
\int_0^\infty \d z\,z^4 f_4(z)=0,
\end{equation}
which means that $\zeta(1)$, which would indicate an unremovable divergence,
does not occur in the summation over $m$.  This is the content of the
proof that the Casimir energy for a dilute dielectric cylinder is finite 
in this order, given by Bordag and Pirozhenko \cite{bordag01}.
We also note that when the divergent part is removed from the $f_2$ integration
we again get zero,
\begin{equation}
\int_0^\infty \d z\left(z^4 f_2(z)-\frac18\right)=0.
\label{ct}
\end{equation}
The suggestion is that this term may be simply omitted as a contact term.
(But see below.)

However, the two logarithmically divergent terms, corresponding to
$f_1$ and $f_3$, give finite contributions, because they are multiplied
by formally zero values of the Riemann zeta function.  The first one may
be regulated by a small change in the power:
\begin{subequations}
\begin{equation}
\lim_{s\to0}\sum_{m=1}^\infty m^{2-s}\int_0^\infty \d z\,z^{4-s}f_1(z)
=\lim_{s\to0}\frac14\zeta(-2+s)\frac1s=-\frac{\zeta(3)}{16\pi^2}.
\label{t1}
\end{equation}
The $f_3$ term gives similarly
\begin{equation}
\lim_{s\to0}\sum_{m=0}^\infty{}'m^{-s}\int_0^\infty \d z\,z^{4-s} f_3(z)
=\zeta'(0)\left(-\frac5{16}\right)=\frac5{32}\ln2\pi.
\label{t2}
\end{equation}
\end{subequations}
Although it would appear that a finite term would emerge from $f_4$, that
term vanishes because remarkably
\begin{equation}
\int_0^\infty \d z\,z^4\,\ln z\,f_4(z)=0.
\end{equation}
The $f_5$ term is completely finite:
\begin{equation}
\sum_{m=1}^\infty \frac1{m^2}\int_0^\infty \d z\,z^4 f_5(z)=\frac{19\pi^2}
{7680}.
\label{t3}
\end{equation}

Following the above prescription, we arrive at the following entirely
finite expression for the pressure on the cylinder:
\begin{eqnarray}
P&=&\frac{(\varepsilon-1)^2}{32\pi^2 a^4}\bigg\{-\frac{\zeta(3)}{16\pi^2}
+\frac5{32}\ln2\pi+\frac{19\pi^2}{7680}\nonumber\\
&&\quad\mbox{}+2\sum_{m=1}^\infty \int_0^\infty \d y\,y^4\left[g_m(y)-
\frac1{2m^2}\sum_{k=1}^5\frac1{m^k}f_k(y/m)\right]\nonumber\\
&&\quad\mbox{}+\int_0^\infty \d y\,y^4\left[g_0(y)-\frac1{16}\frac1{y^4}-
\frac12f_3(y)\right]\bigg\}.
\label{numresult}
\end{eqnarray}
Here in $g_0$ we have subtracted a linearly divergent term, which when
combined with that removed in (\ref{ct}) gives
\begin{equation}
\frac18\sum_{m=0}^\infty{}'\int\d y.
\end{equation}
We regard this, rather cavalierly, as a contact term, which we simply omit.
In the next section we will give the correct treatment of this $f_2$ term. 
All that remains is to do the integrals numerically.  We do so for $m$ from
0 through 4, after which we use the next nonzero term in the uniform
asymptotic expansion,
\begin{equation}
\sum_{m=5}^\infty\int_0^\infty \d z\,z^4\left[\frac1{m^3}f_6(z)+\frac1{m^4}
f_7(z)\right]=-\frac{209}{64512}\sum_{m=5}^\infty\frac1{m^4},
\label{t5}
\end{equation}
because, again, the integral over $f_6$ vanishes.

When all the above is included, to 6 decimal places, we obtain
\begin{eqnarray}P&=&\frac{(\varepsilon-1)^2}{32\pi^2a^4}(
-0.007612+0.287168+0.024417-0.002371-0.000012\nonumber\\
&&\qquad\qquad\mbox{}-0.301590)=0.000000,
\label{zresult}
\end{eqnarray}
where the successive terms come from (\ref{t1}), (\ref{t2}), (\ref{t3}),
the numerical integral over the first 4 subtracted $g_m$s ($m>0$), 
the remainder (\ref{t5}), and 
the numerical integral over the subtracted $g_0$, respectively.
This constitutes a convincing demonstration of the vanishing of the
Casimir pressure in this case. It is similar to the numerical
demonstration \cite{milt-nest-nest} 
of the seemingly coincidental vanishing of the Casimir
energy for a dilute dielectric-diamagnetic cylinder, obtained by 
expanding (\ref{dildieldimag}) to order $\xi^2$.

\subsection{Exponential regulator}
Although the calculation in the previous subsection is quite standard,
and undoubtedly correct, the reader might rightly object that zeta-function
regulation has been employed, and infinite terms simply omitted.  Therefore,
and to make contact with known results, let us insert a regulator to make
all the sums and integrals completely finite.  It would be best, as in 
Ref.~\cite{deraad&milton1981}, to insert such a regulator before rotating the
frequency in the complex plane.  However, this is much more complicated here
than in that reference; and because the expressions here are formally much more
divergent, the regulator adopted there appears insufficient.  It will suffice
for the present purposes to simply insert by hand an exponential regulator
into the expression (\ref{epsilon2}):
\begin{equation}
P_{\rm reg}=\frac{(\varepsilon-1)^2}{16\pi^2a^4}\sum_{m=0}^\infty{}'\int_0^\infty \d y\,
y^4 g_m(y)\e^{-\delta y},
\label{epsilon2reg}
\end{equation}
where $\delta\to0+$ at the end of the calculation.
Then it is easy to repeat the calculation of the previous subsection.
One has only to carry out the sum
\begin{equation}
\sum_{m=1}^\infty\e^{-\delta m z}=\frac1{\e^{\delta z}-1}.
\end{equation}
Then the  $f_1$ term, instead of (\ref{t1}), is
\begin{equation}
\int_0^\infty \d z\,z^4 f_1(z)\frac{\d^2}{\d \delta^2 
z^2}\frac1{e^{\delta z}-1}=
\frac{13\pi}{32\delta^3}-\frac{\zeta(3)}{16\pi^2}.
\end{equation}  
The $f_2$ term has no finite part:
\begin{equation}
-\int_0^\infty \d z\,z^4 f_2(z)\frac{\d}{\d \delta z}\frac1{e^{\delta z}-1}=
-\frac1{16\delta},
\label{f2div}
\end{equation}
where the reader should note that no ad hoc subtraction as in (\ref{ct})
has been employed.  The evaluation of (\ref{f2div}) uses the fact that
\begin{equation}
\int_0^\infty \d z\,z^2\,f_2(z)=0.\label{f22int}
\end{equation}
The $f_3$ term is, instead of (\ref{t2}),
\begin{equation}
\int_0^\infty \d z\,z^4\, f_3(z)\left(\frac1{\e^{\delta z}-1}+\frac12\e^{-\delta
z}\right)=-\frac{315\pi}{8192\delta}+\frac5{32}\ln 2\pi.
\end{equation}
Here we have subtracted a term from the $m=0$ contribution:
\begin{equation}
\int_0^\infty\d y\, y^4\left[g_0(y)-\frac12 f_3(y)\right]\e^{-\delta y}
=-0.301590+\frac1{16\delta}.
\end{equation}
The divergent term here cancels that in (\ref{f2div}), and the finite part is
the value of the last integral in (\ref{numresult}).
Thus we recover exactly the same numerical result (\ref{zresult}) found in 
the previous subsection, plus two divergent terms
\begin{equation}
P_{\rm div}=\frac{(\varepsilon-1)^2}{32\pi^2a^4}
\left(\frac{13\pi^2}{32\delta^3}-\frac{315\pi}{8192\delta}\right).
\end{equation}
The form of the divergences is exactly as expected \cite{bordag01,barton}.
In particular, there is no $1/\delta^2$ divergence, because of the identity
(\ref{f22int}).

\subsection{Interpretation of divergences}
In the previous section we computed divergent contributions to the Casimir
pressure for a dilute cylinder.  For simplicity, we chose an exponential
regulator with a small 
dimensionless parameter $\delta\to0+$.  How do we interpret these
terms?  It is perhaps easiest to imagine that $\delta$ as given in terms of
a proper-time cutoff, $\delta=\tau/a$, $\tau\to 0+$.  Then if we consider the
energy, rather than the pressure, the divergent terms have the form
\begin{equation}
E_{\rm div}=e_3\frac{aL}{\tau^3}+e_1 \frac{L}{a}\frac{1}{\tau}.
\end{equation}
Here $L$ is the (large) length of the cylinder. 
 Thus, the leading divergence
corresponds to an energy term proportional to the surface of the cylinder,
and it therefore appears sensible to absorb it into a renormalized surface
energy which enters into a phenomenological description of the material system.
(Such arguments are familiar, dating back to \cite{milton1980}, and recently
vigorously revived in \cite{bv}.)
The $1/\tau$ divergence is more problematic.  It is proportional to the
ratio of the length to the diameter of the cylinder, so it seems likely that
this would be interpretable as an energy term referring to the shape of the
body.  If one could compute the Casimir energy of an extremely elongated
ellipsoid, we would expect an energy term proportional to the ratio of
curvatures.  (Of course, a cylinder has zero curvature.)
  This appears to be exactly of the form of a surface integral \cite{Candelas}
\begin{equation}
\int \d S\,\kappa_1\kappa_2,
\end{equation}
in terms of the principal curvatures $\kappa_i$, $i=1,2$.  Such terms are
well known not to contribute to the observable energy.
Had a divergent term proportional to $\delta^{-2}$ appeared in the pressure,
it would have implied a divergent energy of the form $e_2(\ln a)/\tau^2$,
which would have been impossible to remove.  (For the dielectric sphere
the situation is simpler, in that divergences are all associated with
positive powers of the sphere's radius \cite{ddsph}.)

In any case, although the structure of the divergences is universal,
the coefficients of those divergences depend in detail upon the particular
regularization scheme adopted.  In contrast, the term proportional to 
$(\varepsilon-1)^2/a^2$ is unique.  Thus, of course, it could not have been
any other than that zero value given by the van der Waals calculations 
\cite{milt-nest-nest,milonni,romeopc}.  

The nature of divergences in such Casimir calculations is still under
active study \cite{jaffe,fulling,casrev}.  The universality of the
finite Casimir term makes it hard not to think it has some real significance.
As an example of how subtle interpretation of divergences can be, we recall
that it has now been proved that the total Casimir energy for electromagnetic
modes interior and exterior to an arbitrarily shaped smooth infinitesimally
thin closed perfectly conducting
surface is finite \cite{graf}.  This is hard to reconcile
with the existence of local divergences in the energy density near the
surface proportional to $(\kappa_1-\kappa_2)^2$.  Presumably, these
divergences belong to the surface itself and have nothing to do with the
global Casimir energy \cite{fulling,casrev}.  But the open questions are
profound and challenging.

\section{Conclusions}
Since the beginning of the subject, the identity of the Casimir force
with van der Waals forces between individual molecules has been evident
\cite{casimirpolder,casimir}.  It is essentially just a change of perspective
from action at a distance to local field fluctuations.  So it was no
surprise that the retarded dispersion force between molecules, the 
Casimir-Polder force, could be derived from the Lifshitz force between
parallel dielectric surfaces \cite{lifshitz,schdermil}.
However, the identity is not really that trivial, because both the
van der Waals and the Casimir energies contain divergent contributions.
This is particularly crucial when one is considering the self-energy
of a single body rather than the energy of interaction of distinct bodies.
Thus it was nontrivial when it was proved that the Casimir energy of 
a dilute dielectric sphere \cite{ddsph} coincided with that obtained
by summing the van der Waals energies of the constituent molecules
\cite{milton1998}.

When it was shortly thereafter discovered that the sum of van der Waals forces
vanished for a dielectric cylinder \cite{romeopc,milt-nest-nest} it was 
universally believed that the corresponding Casimir energy, in the dilute
approximation, must also vanish. 
This result was verified by a perturbative calculation \cite{barton}.
 Proving this by a full Casimir calculation turned out to be extraordinarily
difficult.  This paper is the  result of a five-year-long effort.  
It should dispell any lingering doubts about the meaning of the Casimir
force.  The importance of this finding is impossible to evaluate at this
point; a zero value suggests some underlying symmetry, which is certainly
far from apparent.  It probably has technological implications, for
example in the physics of nanotubes, which will be explored in a subsequent
publication. 
\begin{ack}

We thank the US Department of Energy for partial support of this
research.  We acknowledge numerous communications with August Romeo,
and many helpful conversations with K. V. Shajesh.

\end{ack}

\end{document}